# A fast new catadioptric design for fiber-fed spectrographs

Will Saunders[1]

Australian Astronomical Observatory, Sydney NSW 1710, Australia

## ABSTRACT

The next generation of massively multiplexed multi-object spectrographs (DESpec, SUMIRE, BigBOSS, 4MOST, HECTOR) demand fast, efficient and affordable spectrographs, with higher resolutions ($R$ = 3000-5000) than current designs. Beam-size is a (relatively) free parameter in the design, but the properties of VPH gratings are such that, for fixed resolution and wavelength coverage, the effect on beam-size on overall VPH efficiency is very small. For all-transmissive cameras, this suggests modest beam-sizes (say 80-150mm) to minimize costs; while for catadioptric (Schmidt-type) cameras, much larger beam-sizes (say 250mm+) are preferred to improve image quality and to minimize obstruction losses. Schmidt designs have benefits in terms of image quality, camera speed and scattered light performance, and recent advances such as MRF technology mean that the required aspherics are no longer a prohibitive cost or risk.

The main objections to traditional Schmidt designs are the inaccessibility of the detector package, and the loss in throughput caused by it being in the beam. With expected count rates and current read-noise technology, the gain in camera speed allowed by Schmidt optics largely compensates for the additional obstruction losses. However, future advances in readout technology may erase most of this compensation.

A new Schmidt/Maksutov-derived design is presented, which differs from previous designs in having the detector package outside the camera, and adjacent to the spectrograph pupil. The telescope pupil already contains a hole at its center, because of the obstruction from the telescope top-end. With a 250mm beam, it is possible to largely hide a 6cm × 6cm detector package and its dewar within this hole. This means that the design achieves a very high efficiency, competitive with transmissive designs. The optics are excellent, as least as good as classic Schmidt designs, allowing F/1.25 or even faster cameras. The principal hardware has been costed at $300K per arm, making the design affordable.

## 1. INTRODUCTION

Volume Phase Holographic (VPH) gratings consist of a thin layer of Dichromated Gelatin (DCG) sandwiched between glass substrates. The DCG contains a sinusoidal variation in refractive index, causing constructive interference at a wavelength-dependent angle, and hence acts as a disperser. VPH gratings have multiple advantages over traditional ruled reflection gratings:

- they are much more efficient, especially near the blaze angle;
- they are tuneable, i.e. they can be made to exactly the required size, dispersion and blaze wavelength;
- they are much cheaper, and there are multiple vendors;
- they are available in large sizes;
- they have lower scattered light;
- because they work in transmission, the required pupil relief is smaller than for reflection gratings, so (for given beam-size and camera speed and angles) the other optics can be smaller.

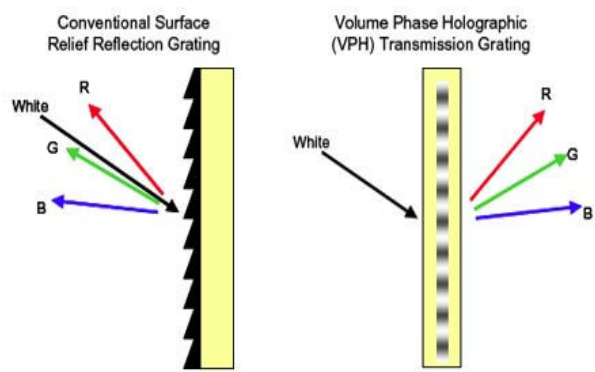

**Figure 1. Schematic representation of VPH vs reflection gratings**

[1] will@aao.gov.au

Because of these advantages, VPH gratings are now universal in low-dispersion (say R<10,000) spectrograph design. For a fixed-format spectrograph (i.e. with fixed dispersion and wavelength coverage), it is sensible to include the VPH efficiency characteristics explicitly in the spectrograph design, to maximize overall efficiency. However, while reflection gratings have just two primary degrees of freedom (ruling spacing and blaze angle), VPH gratings have three (fringe spacing, DCG thickness, DCG index variation), all have strong effects on efficiency envelope. This extra degree of freedom makes the calculation of optimal parameters rather complex, and the next few sections represent an attempt to signpost the optimal procedure.

In any design, most of the spectrograph parameters are fixed from the outset. They would normally include

- External parameters such as fiber diameter, input beam speed, and (with only a few discrete choices) the number and size of pixels in the spectral direction on the detector.
- The hard requirements of wavelength coverage and resolution.
- The softer requirements of efficiency envelope, imaging quality and uniformity, and cost.

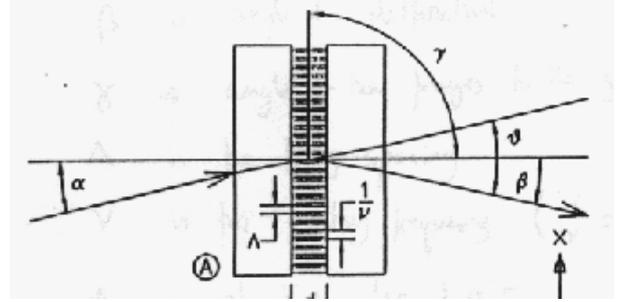

Figure 2. VPH grating layout and parameters. For Littrow configuration and unslanted fringes, $\alpha=\beta=\varphi$ and $\gamma=90°$.

For each arm, the required dispersion $\delta\lambda$ (wavelength per pixel) is given by $\delta\lambda = \Delta\lambda / N$, where $\Delta\lambda$ is the required wavelength coverage and $N$ the number of spectral pixels on the detector. The required FWHM resolution $r$ (in pixels) is then given by the required dimensionless resolution $R$ at the wavelength of interest:

$$R = \lambda / (r\, \delta\lambda) \qquad (1)$$

The required demagnification $\mu$ is then determined by the fiber diameter $d$ and the pixel size $p$:

$$\mu = F_{cam} / F_{coll} \approx 6/5\, r\, p\, /\, d \qquad (2a, 2b)$$

Where $F_{cam}$ and $F_{coll}$ are the camera and collimator speeds. The factor 6/5 arises because, for realistic aberrations (i.e. small but non-negligible compared to the projected fiber size) the FWHM of the aberrated circular image of the fiber is ~5/6 times its unaberrated diameter [1,2].

Eliminating $r$ from (1) and (2b) gives

$$R \approx 6/5\, (\lambda/\delta\lambda)\, (p/d)\, (F_{coll}/F_{cam}) \qquad (3)$$

The collimator speed $F_{coll}$ is determined by the input beam speed[2], so for given wavelength coverage and $R$, $p$, and $d$, the camera speed $F_{cam}$ is then fixed also.

To find the actual grating parameters (grating angle and line density) we need to achieve the required resolution and wavelength coverage, we need the grating equation and its derivative. For Littrow configuration[3] and first order use, the grating equation is

$$\lambda = 2 \sin\alpha\, \Lambda = 2\, n\, \sin\alpha_2\, \Lambda \qquad (4a, 4b)$$

where $\lambda$ is the central wavelength, $\alpha$ is the grating angle in air, $\Lambda$ is the line spacing, $n$ is the average refractive index of the DCG, and $\alpha_2$ is the grating angle in the DCG (given by $\sin\alpha = n \sin\alpha_2$). The derivitive of the grating equation with respect to exit angle is

$$\delta\lambda = \cos\alpha\, \Lambda\, \delta\alpha = \cos\alpha\, \Lambda\, p / (B\, F_{cam}) \qquad (5a, 5b)$$

where $B$ is the beam diameter (so $B\, F_{cam}$ is the focal length of the camera).

---

[2] Usually 5-10% faster than the fiber input beam speed, to allow for Focal Ratio Degradation within the fibers.
[3] A small deviation from Littrow allows elimination of a troublesome ghost, of light reflected back from the detector and recombined by the grating in -1R order [1]

Assuming the pixel size, dispersion and coverage are all fixed by the requirements, then we have two equations (4a and 5b) for the three unknowns $\alpha$, $\Lambda$ and $B$. Therefore, we have freedom to choose one, which we take to be $B$, before solving for the others. We are far from the diffraction-limited regime (i.e. $B \gg R\lambda$), so we have considerable freedom in choosing $B$. For any given $B$, the resulting $\alpha$ and $\Lambda$ can then be determined, and the DCG thickness and index modulation then tuned for best efficiency. So our choice of $B$ has not only large and direct implications for image quality and cost, but also indirect but potentially large effects on VPH efficiency.

## 2. VPH CHARACTERISTICS

AAOmega [1] was the first general-purpose low-dispersion spectrograph to be designed around VPH gratings. The intention was to provide $R \sim 1500-10{,}000$ at all wavelengths 370nm-950nm. Figure 3 shows the measured efficiency profiles for all the AAOmega gratings, (all supplied by Richard Rallison at Ralcon). This plot illustrates various properties of VPH gratings:

- Up to 90% efficient at peak.
- Narrower bandwidth than reflection gratings.
- Peak efficiency drops at higher dispersions because $s$ and $p$ polarisations cannot be simultaneously optimised.
- Peak efficiency also drops at lower dispersions because of losses to 0$^{th}$ order.

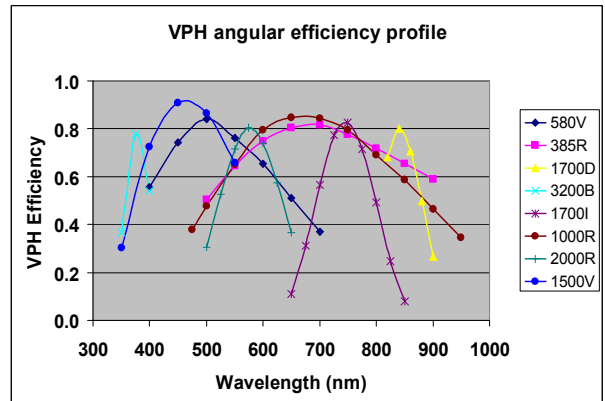

**Figure 3. Efficiency profiles of all AAOmega gratings**

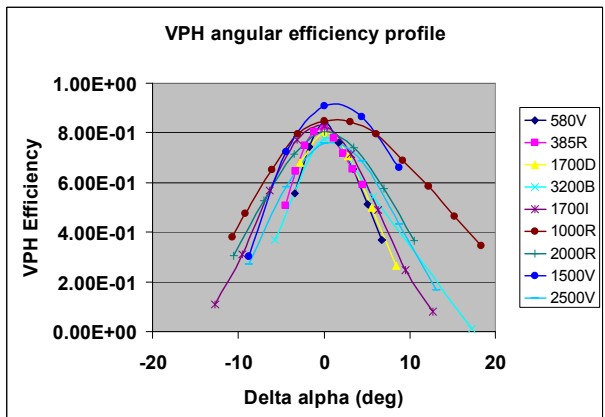

**Figure 4. Angular efficiency profile for AAOmega gratings**

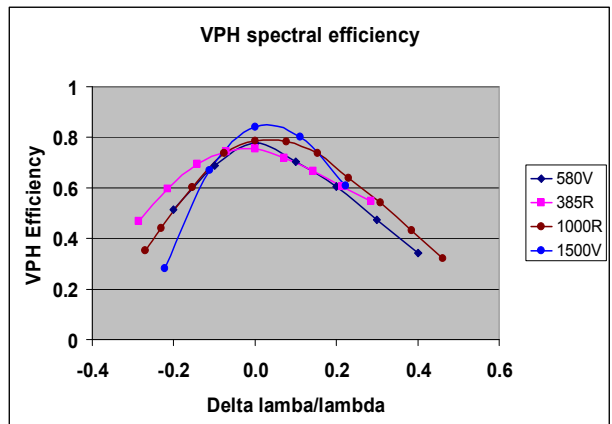

**Figure 5. Spectral efficiencies for gratings with grating angles ~8° and ~20°**

Figure 4 shows the angular efficiency envelope for each grating; that is the x-axis shows the angular deviation from the blaze angle (in air), on exit from the grating. Most of the gratings have very rather similar angular efficiency profiles. If all gratings had the same angular efficiency profile, it would favour large spectrograph beam-sizes, since (for fixed spectral coverage, detector size and camera speed), the angular spread declines as the camera length increases. However, two gratings (1000R and 1500V) stand out from the rest, as having both higher peaks and broader bandwidths. The blaze grating angles in both cases are about 20°. Their *angular* bandwidths are so good, that their *spectral* bandwidths are about as good as the much lower dispersion gratings (385R and 580V) covering the same wavelength ranges (Figure 5). The next few sections are devoted to understanding this effect and its implications.

## 3. VPH PEAK EFFICIENCY

The efficiency characteristics of VPH gratings were investigated by Kogelnik [3], while our treatment and notation follow that of Baldry *et al*. [4]. Kogelnik derived an approximate peak efficiency (the Kogelnik approximation)

$$\eta \approx \tfrac{1}{2} \sin^2(\pi \Delta n\, d / \lambda \cos\alpha_2) + \tfrac{1}{2} \sin^2(\pi \Delta n\, d \cos 2\alpha_2 / \lambda \cos\alpha_2) \qquad (6)$$

where $\Delta n$ is the index modulation, $d$ the DCG thickness, $\alpha_2$ the grating angle in DCG, and the two terms refer to $s$ and $p$ polarizations respectively. For small angles, close to 100% peak efficiency is possible, but at larger angles the cos $(2\alpha_{2b})$ term means the $s$ and $p$ polarizations cannot be simultaneously optimized. The maximum efficiency is achieved, for unpolarised light, when

$$\Delta n\, d \approx \tfrac{1}{2} \sec\alpha_2\, \lambda \qquad (7)$$

as a compromise between the $s$ and $p$ polarizations The approximation is valid as long as

$$\rho = \Delta n / (n \sin^2\alpha) > \sim 10 \qquad (8)$$

Typically $n \sim 1.3$, and $\Delta n$ can be ~0.1 or more. Large $\Delta n$ is in general desirable, see below)[4], so the Kogelnik approximation is always valid for grating angles $\alpha >\sim 25°$. The regime of validity is most easily understood by combining (4a), (7) and (8) to obtain

$$d \tan\alpha_2 = \rho/4\, \Lambda \qquad (9)$$

This means that the angle of incidence and the depth of the grating, must be large enough for undeviated rays to pass through multiple fringes before they exit. For smaller thicknesses (and especially, $\rho < \sim 4$) a significant fraction of the light does indeed pass straight through the grating (zeroth order losses). So to avoid 0th order losses, there is a minimum DCG thickness $d$, and hence (from (7)) a maximum index modulation $\Delta n$. The $0^{th}$ order losses are very crudely $1/2\rho^2$ (see below), so values of $\rho \sim$ few do not lead to excessive losses.

## 4. VPH BANDWIDTH

The half-power bandwidth was crudely derived by Kogelnik, as

$$\Delta\lambda_{\text{eff}}/\lambda \sim \Lambda \cot\alpha_2 / d \qquad (10)$$

We can rewrite this in angular terms, using eqns (5a) and (7), to get

$$\Delta\alpha \sim 2\, \Delta n \cos^2\alpha_2 / \sin\alpha_2 \cos\alpha \sim 2\, \Delta n / \sin\alpha_2 \qquad (11)$$

So for grating angles $>\sim 25°$, the efficiency is maximised by fixing $\Delta n$ to be as large as possible (say 0.1), then selecting $d$ from equation (7). The formula shows the decline in angular bandwidth with grating angle, as observed. At smaller grating angles we can combine (10) and (8) to get the simple result

$$\Delta\lambda_{\text{eff}}/\lambda \sim 4/\rho \qquad (12)$$

This shows immediately, that (a) we cannot have gratings obeying the Kogelnik approximation ($\rho > \sim 10$) broad enough to cover the entire 370-950nm band in one or even two arms (which requires $\Delta\lambda_{\text{eff}}/\lambda >\sim \tfrac{1}{2}$), and that (b) lowering $\rho$ broadens the bandwidth. However, once we have $\rho << 10$, the Kogelnik analysis is no longer valid and equation (12) is no longer quantitatively useful.

## 5. VPH EFFICIENCY PROFILES AND SPECTROGRAPH DESIGN

We have therefore investigated the bandwidth directly, using the GSOLVER coupled-wave software to model a set of gratings covering a wide range of line densities and DCG thicknesses. We assumed a required wavelength coverage of 600nm-1000nm with a single 6cm × 6cm detector, with camera focal lengths of 300mm, 240mm, 200mm, 150mm, 120mm, covering the likely range of possibilities. The resulting line densities and wavelength coverage are shown in Table 1. For each line density, we modeled a range of thicknesses and index modulations, all with $\Delta n \times d = 0.375\mu m$, and spanning the useful range in $\Delta n$. All gratings were assumed to have $n = 1.4$.

| Camera focal length $f_{\text{cam}} = B\, F_{\text{cam}}$ | Line density $\nu = 1/\Lambda$ | Grating angle in air $\alpha$ | Grating angle in DCG $\alpha_2$ |
|---|---|---|---|
| 300mm | 490/mm | 11.3° | 7.5° |
| 240mm | 606/mm | 14.0° | 9.3° |
| 200mm | 718/mm | 16.7° | 11.0° |
| 150mm | 928/mm | 21.8° | 14.3° |
| 120mm | 1118/mm | 31.0° | 20.0° |

**Table 1. VPH grating parameters explored for modelling**

---
[4] $n$ and $\Delta n$ are anti-correlated (e.g. [4]), with $n$ varying from 1.54 to 1.25 as $\Delta n$ varies from 0 to 0.1, but this effect is not taken into account in this paper. The value of $n$ has only a weak (~1%) effect on VPH properties, but in the sense of favouring larger beam-sizes.

Figure 6 shows the results for the 606/mm gratings. The peak efficiency drops as the thickness decreases (because of $0^{th}$ order losses), but bandwidth increases greatly. The average efficiency has a maximum at 84%, for a DCG thickness 7.5μm, corresponding in this case to $\rho \sim$ few. The dependence of average efficiency on DCG thickness is very weak in the regime $5 \leq d \leq 10$.

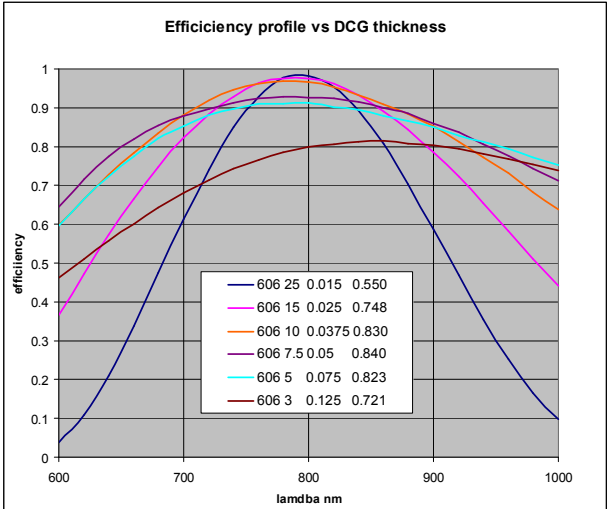

Figure 6. Efficiency profile for a set of VPH gratings. Numbers are lines/mm, *d*, Δ*n*, and average efficiency.

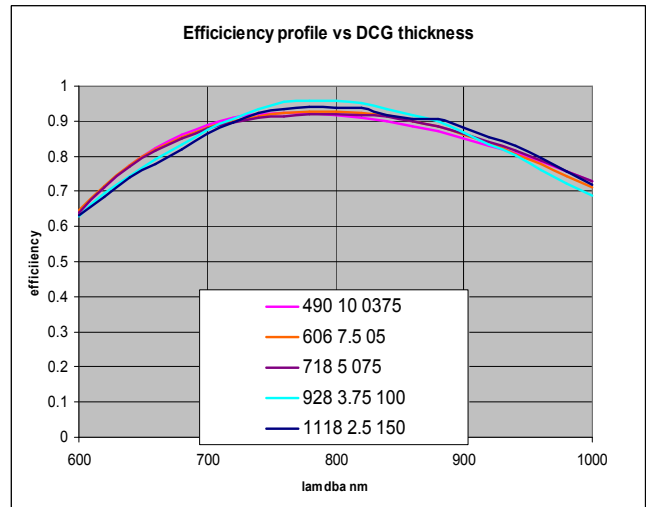

Figure 7. Efficiency profile for the most efficient grating in each set. Numbers are lines/mm, *d*, and Δ*n*,. Efficiency is always 84%.

This exercise was repeated for all the line densities in Table 1, and the set of 'best' gratings is shown in Figure 7. The curves are strikingly similar, despite the huge differences in thickness, index modulation, grating angle and line density. In every case, the best average efficiency was very close to 84%. The coarser gratings have thicker DCG, and hence decreased angular bandwidth, but this is almost exactly compensated by the smaller field angles they require for the same spectral coverage. There is a weak effect that the coarser gratings have larger bandwidths and lower peak efficiencies (because of $0^{th}$ order losses).

The main implication for spectrograph design is that beam-size does not, after all, have any first-order effect on overall efficiency. The designer is free to find the smallest camera length (right down to ~100mm) consistent with the diffraction limit, adequate geometric image quality and (for reflective designs) detector obstruction losses.

## 6. READ NOISE AND CAMERA SPEED

Various factors drive the designs into a regime where read-noise is significant. Figure 8 shows the night sky spectrum for a dark mid-latitude site, in units of erg/μm/s/m$^2$/″$^2$. When translated into photons rather than ergs, the continuum is nearly uniform between 350nm and 950nm. So at fixed dispersion (Δ$\lambda$/pixel), the continuum sky is just as dark in the far red as in the UV. In general, the targets are faint objects in natural seeing, contributing fewer photons than the sky. All the proposed instruments are designed to work between the OH lines, driving the resolution to large values, R=3000-5000. Some of the instruments (e.g. BigBOSS, DESpec) are proposing very short total integration times per field, which when combined with the need for a minimum of three frames/field (for cosmic ray removal), imply very short individual exposures of 10mins or less. With such dark sky, faint objects, high resolutions, and short exposures, there is an obvious risk of significant read-noise and/or read-time penalty to the overall survey efficiency. The actual count rate is given by

$$n = \eta \, s \, \pi D^2/4 \, \pi a^2/4 \, \delta\lambda \, / d_p \quad (10)$$

where $n$ is the count rate in photons/pixel/s, $\eta$ is the system efficiency, $s$ is the sky flux (photons per unit area per unit solid angle per unit wavelength), $D$ is the telescope diameter, $a$ the angular fiber aperture diameter, $\delta\lambda$ the dispersion (wavelength per pixel), $d_p$ is the projected fiber size in pixels on the detector. Including explicitly the spectral resolution $R$, and the telescope, collimator and camera speeds $F_{tel}$, $F_{coll}$ and $F_{cam}$, and solving for $d_p$ from (0) gives, in time $t$, a number of photons per pixel of

$$N = \eta \, s \, (\pi/4)^2 \, (5/6) \, 0.205^2 \, p^2 \, \lambda/R \, t \, (F_{coll}/F_{tel})^2 / F_{cam}^2 \quad (11)$$

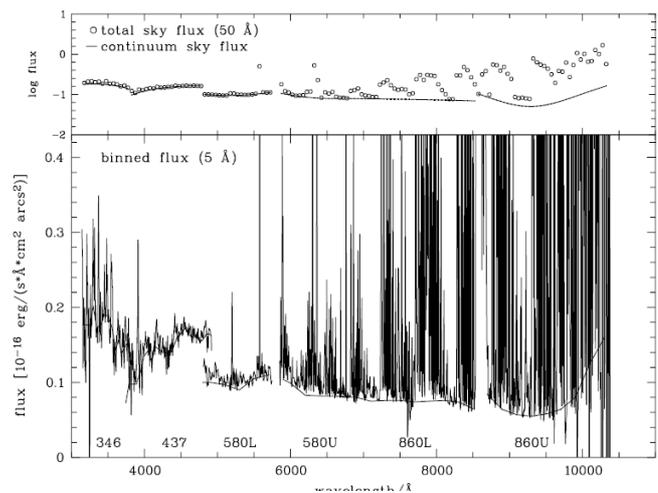

Figure 8. Night sky spectrum, for Cerro Paranal [5].

For values relevant to DESpec ($\eta = 0.25$, $s = 240$ $v/s/\mu m/m^2/''^2$, $p = 15\mu m$, $\lambda = 0.8\mu m$, $R = 3000$, $t = 600s$, $F_{coll} = 2.75$, $F_{tel} = 2.9$), this gives just $\sim 50/F_{cam}^2$ detected sky continuum photons per pixel in the peak spatial pixel of the spectrum, and less in the wings. This means that the read-noise will make a significant contribution to the total noise, even for state-of-the-art read-noises of $\sim 2.5e^-$. The calculated efficiency hit as a function of camera speed and read noise (calculated from the increased observing time needed to reach the same S/N) is shown in Figure 9 and Table 2. The read noise can be decreased by slower read-out speeds, but this increases the efficiency loss (additional to that in Figure 9 and Table 2) from the time lost physically reading out the detector. Even to reach a read-noise of $2.5e^-$ typically takes read-out times of 50-100s, significant compared to the integration times per frame, and so representing a significant additional efficiency hit. So for current read-out technologies, it makes sense to always use the fastest cameras compatible with adequate sampling of the PSF.

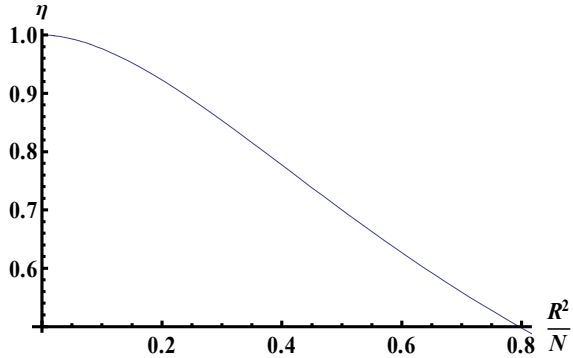

Figure 9. Read-noise efficiency penalty (equivalent throughput for the same S/N), as a function of read-noise and the number of counts in the peak pixel.

| Pixel size/camera speed (Speed for 15μm pixels) | 12μm (F/1.25) | 10μm (F/1.5) | 7.5μm (F/2.0) |
|---|---|---|---|
| RN/pixel | | | |
| 1.0 e$^-$ | 0.997 | 0.994 | 0.984 |
| 1.5 e$^-$ | 0.987 | 0.976 | 0.936 |
| 2.0 e$^-$ | 0.965 | 0.936 | 0.840 |
| 2.5 e$^-$ | 0.923 | 0.868 | 0.702 |
| 3.0 e$^-$ | 0.868 | 0.775 | 0.549 |

Table 2. Read-noise efficiency penalty (calculated as the equivalent throughput giving the same S/N) as a function of camera speed, pixel size and read noise. The hit from read-out *time* is not included.

The effect is large enough to make catadioptric cameras competitive in survey efficiency, just because of their greater speeds. However, in the future, the problem is likely to be greatly alleviated by the digital filtering techniques of e.g. [6], offering 1-2e$^-$ read-noises with realistic (50-100kHz) read-out rates. These techniques, if proven and used, would restore the efficiency benefits of transmissive cameras (see below).

## 7. TRANSMISSIVE VS REFLECTIVE OPTICS

The collimator design is both easier and has less impact on overall performance than the camera design. For fiber-fed collimators, speeds are typically F/2.5 - F/4, giving a choice of on-axis-reflective, or off-axis-reflective transmissive designs. The first of these has a throughput penalty from the slit being in the beam, but easier optics. The throughput loss is greatly reduced if a fold-mirror with a minimal slot is employed (as for VIRUS [7]). However, probably more significant for fiber-fed spectrographs is the ease of back-illumination of the fibers (to allow closed-loop metrology for multi-object spectroscopy), preferably while reading out the detectors (to save the time otherwise spent repositioning between fields). For pick-and-place robotic positioners with multiple field-plates (such as 2dF), this can be done with a physical slit-exchange mechanism, but otherwise, a completely light-tight fold-mirror or shutter must be introduced in front of the slit. This seems plausible for on-axis reflective designs with a fold-mirror, or off-axis reflective or transmissive designs, but seems very difficult for an on-axis reflective design without a fold-mirror.

For the cameras, every spectrograph design faces the fundamental choice of using an all-transmissive, or catadioptric (reflective + transmissive corrector) design. Reflective designs are typically less efficient, because the detector is necessarily within the beam, causing obstruction losses. The detector is also very inaccessible, being completely within the camera. This latter problem can be solved (at some additional cost in obstruction losses) by the 'lollipop' dewar design from SpecInst, which allows the camera to be simply bench mounted without cooling or evacuation.

Catadioptric designs can be faster, they have more uniform image quality (especially in the blue), and they have fewer surfaces and thus less scattering. Scattering is very important for R~3000-5000 work between OH lines, because the gaps are typically only a few resolution elements wide, so easily swamped by the wings of the adjacent OH lines. The choice between transmissive and reflective cameras requires a detailed study, including all these effects.

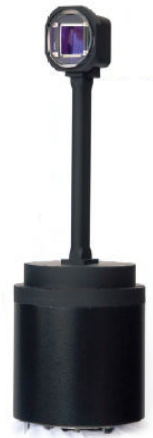

**Figure 10. 'Lollipop' dewar from SpecInst. The example shown is for a 3cm × 3cm detector only, but 6cm × 6cm versions are feasible**

The issue of read noise is potentially a show-stopper. F/1.5 cameras (very fast even in the red, and very difficult in the blue) will have an efficiency hit compared with F/1.25 cameras of 5-10%, and slower cameras become worse very quickly. One intriguing solution to the read-noise problem is to build much slower transmissive cameras, to give FWHM resolutions close to 5 pixels (for DESpec, a camera speed of around F/2.5), and use 2×2 binning in the CCD readout. The benefits are easy optics, and the possibility to use very small beam sizes (say 50-100mm), while still have suitable camera lengths (120-300mm) for using VPH gratings. The cost is in the larger number of spectrographs and total CCD area needed. With 5-pixel FWHM), we can just about reach R~ 0.3nm while still covering the whole optical range (say 370-1000nm) in 3 arms. This might be an interesting modification of the BigBOSS spectrograph design, which originally was already close to this design, with a ~80mm beam and F/2 cameras.

## 8. AN EXISTING TRANSMISSIVE CAMERA DESIGN

An excellent example of a suitable transmissive design , for comparison with the new design presented below, is the JHU adaptation of the SDSS design for WFMOS [8] (Figure 11). This is 2-armed, with F/1.5 cameras and 4K × 4K detectors. The VPH gratings are sandwiched between prisms but this is not a necessity. Figure 12 shows the throughput. Excellent peak efficiency is achieved, but there is a very bad trough at 620-720nm, due to the VPH gratings being stretched beyond the bandpass shown in Figure 7. For DESpec, adequate coverage 400-950nm could be achieved, with efficiency > ~60% for 450-950nm, and resolution R~3500 at the red end of each arm.

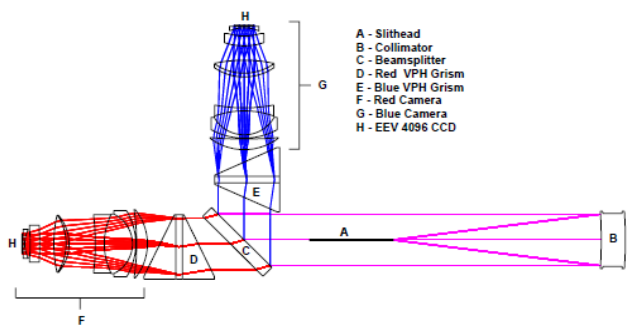

**Figure 11. Layout for the JHU F/1.5 transmissive design. Beam-size is 159mm.**

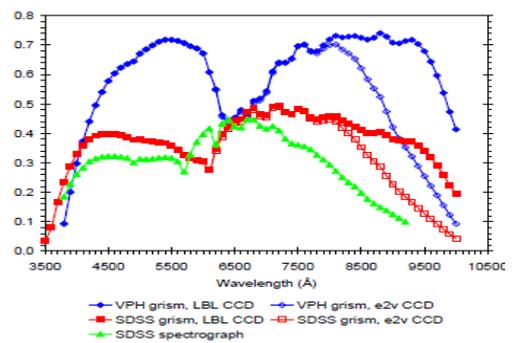

**Figure 12. Throughput for the JHU WFMOS F/1.5 design (upper blue line) [8].**

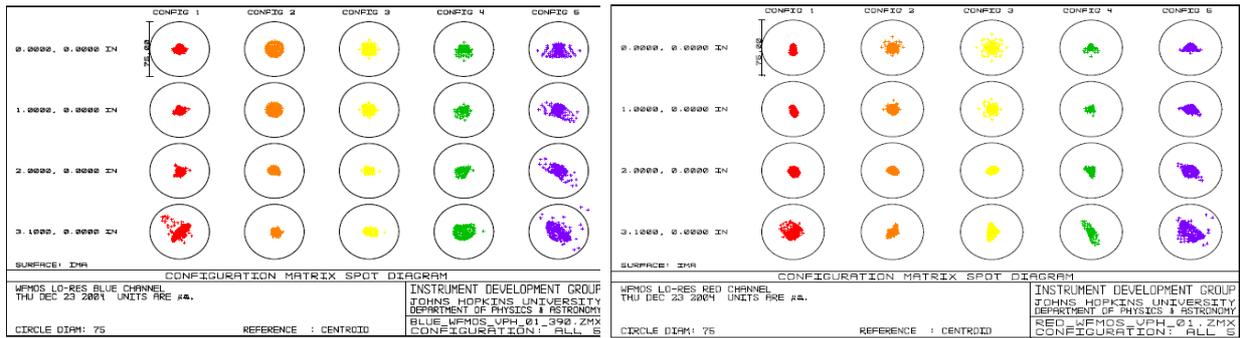

**Figure 13. Spot diagrams for blue (390-655nm, left) and red (600-1000nm right) arms of the JHU F/1.5 transmissive design. Circle size is 75μm. 1.8″ apertures project to ~ 55μm on the detectors.**

The image quality is reasonable (Figure 13), though clearly non-uniform along the slit at the red and blue end of both cameras, potentially compromising sky subtraction[5].

## 9. A NEW REFLECTIVE CAMERA DESIGN

The biggest objection to reflective cameras is the obstruction losses caused by the detector being within the beam. However, this effect is reduced for any system fed by a wide-field telescope, by the fact that the telescope beam itself contains a large hole (typically a third of the pupil diameter), caused by the obstruction of the top end. For a Schmidt-style camera, there is a partial shadowing between these obstruction losses, depending on the field angle. If the detector was closer to the spectrograph pupil (which is usually at the grating), then this shadowing could be improved greatly, making reflective designs competitive in throughput with transmissive designs, but hopefully keeping their other advantages[6].

Maksutov cameras already have the focal plane much closer to the pupil than for a classical Schmidt. Figure 14 shows a Maksutov variant which is believed to be new, where the light passes through the Maksutov corrector in the camera twice. This allows the detector to be placed entirely outside the camera, and as close to the pupil as the detector package allows[7]. To hide the detector package entirely within the footprint of the top-end obstruction, requires a large beam size, say 250mm for a 6cm × 6cm detector.

Both correctors are N-FK5/LF5 doublets, with one external face aspheric, as is the internal interface (but requiring very low surface accuracy). The N-LAK33 field flattener is also aspheric on one surface. Image quality or speed can be improved further if more exotic glasses (PK51A or FK51A) are used in the correctors, or if the field flattener is made a doublet.

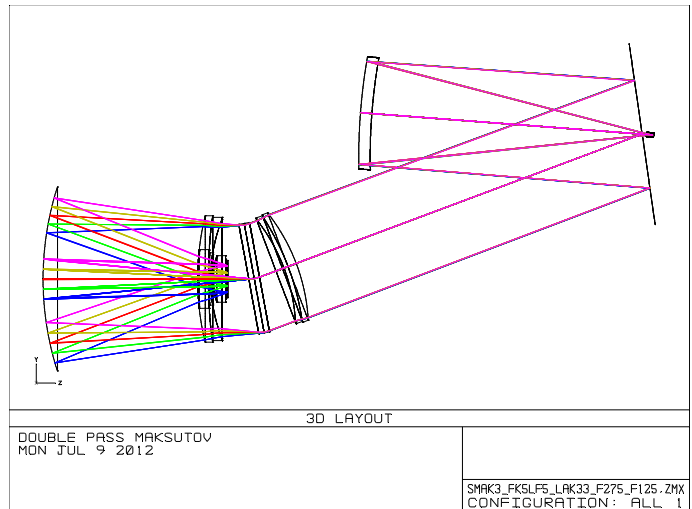

**Figure 14. Layout for red arm of the double-pass Maksutov F/1.25 design. The fold-mirror is to allow easier back-illumination of the fibers. Dichroic for blue arm would be before collimator corrector.**

This image quality is as good as classical Schmidts, and an F/1.25 camera gives maximum rms spot sizes 10μm in the blue and 8μm in the red (Figure 15). This is as good (relative to the projected fiber size) as the F/1.5 transmissive cameras described above.

---

[5] PCA-type sky subtraction can successfully deal with a varying PSF along the slit. However, the quality of the resulting sky subtraction, still correlates with the PSF uniformity. The level of sky subtraction accuracy, and hence PSF uniformity, required for a given project, can only be determined by detailed simulation.

[6] For fiber-fed spectrographs, this shadowing is always imperfect, because FRD tends to fill in the central hole in the pupil.

[7] In principal, the grating could even be at the pupil, with the grating made with a central hole to accommodate it

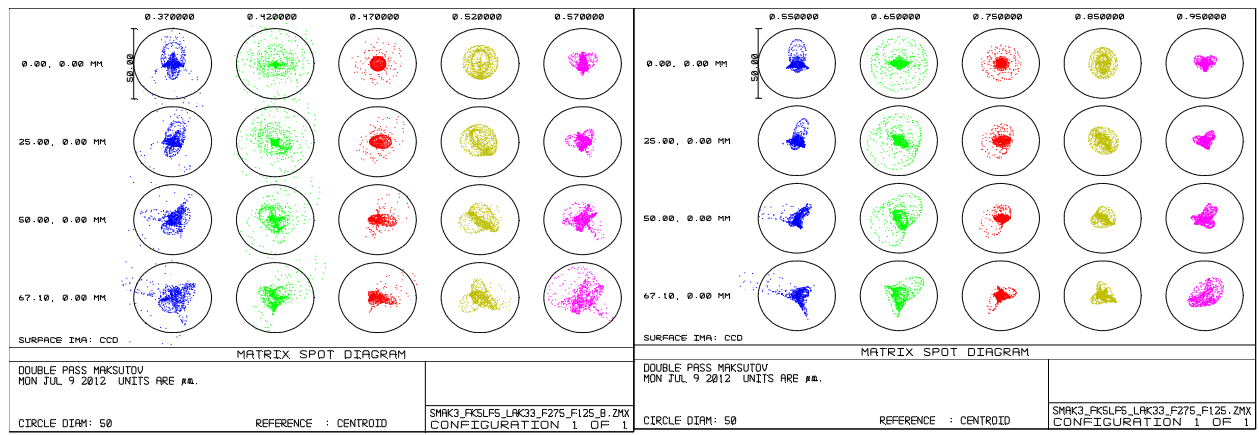

**Figure 14. Spot diagrams for blue (370-570nm, left) and red (550-950nm, right) arms of the double-pass F/1.25 Maksutov design. Circle size is 50μm, projected fiber size for 1.8″ apertures is 45μm.**

Throughput has been calculated and is shown in Figure 16. The calculation assumes 10% vignetting loss (corresponding to the central hole being ~50% filled in by FRD, and with a few mm width assumed for the hole in the collimator fold-mirror for the slit). Grating parameters were derived from GSOLVER for 770/mm (blue) and 490/mm gratings (red), CCD efficiencies were taken from E2V (blue) and DECam (red). The overall peak efficiency is 63% (blue) and 69% (red), with a throughput of 50% or more for 410-940nm.

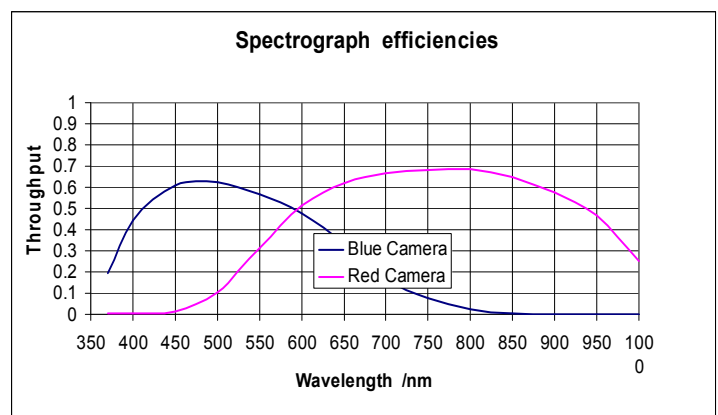

**Figure 16. Estimated throughput for each arm of the double-pass Maksutov design. Dichroic efficiency is not included.**

## REFERENCES

[1] Saunders, W., *et al.,* "AAOmega: a scientific and optical overview," Proc. SPIE 5492 389 (2004).
[2] Sharp, R., *et al.,* "Performance of AAOmega: the AAT multi-purpose fibre-fed spectrograph," Proc SPIE 6269E 14 (2006).
[3] Kogelnik, H., "Coupled Wave Theory for Thick Hologram Gratings," Bell Syst. Tech. J., 48 2909 (1969).
[4] Baldry, I.K., Bland-Hawthorn, J., and Robertson, J. G., "Volume Phase Holographic Gratings: Polarisation Effects and Diffraction Efficiency," Publications of the Astronomical Society of the Pacific, 116 403–414 (2004).
[5] Hanuschik R., W., "A flux-calibrated, high-resolution atlas of optical sky emission from UVES," Astron. Astrophys. 407, 1157-1164 (2003).
[6] Estrada, J.C., "Subelectron readout noise in CCDs," Proc. SPIE 8453 [8453-50] (2012).
[7] Tufts, *et al.*, "VIRUS-P: camera design and performance," Proc. SPIE, 7021-10 (2008).
[8] Smee, S. A., Barkhouser, R. H., Glazebrook, K., "Design of a multi-object, high throughput, low resolution fiber spectrograph for WFMOS," Proc. SPIE 6269 62692I (2006).